\documentclass[a4paper,fleqn,usenatbib]{mnras}

\usepackage{newtxtext,newtxmath}
\usepackage[T1]{fontenc}
\usepackage{ae,aecompl}

\usepackage{graphicx}	% Including figure files
\usepackage{amsmath}	% Advanced maths commands
\usepackage{amssymb}	% Extra maths symbols
\usepackage{epsfig}

\usepackage[usenames,dvipsnames]{color}
\definecolor{purple}{rgb}{0.5,0,0.5}

\newcommand*\dif{\mathop{}\!\mathrm{d}}

\title[Mass Functions, Main Sequence, \& Mergers]{Reconciling Mass Functions with the Star-Forming Main Sequence Via Mergers}

\author[Steinhardt, Yurk, \& Capak]{
Charles. L. Steinhardt$^{1,2,3}$, 
Dominic Yurk$^{2,3}$, 
Peter Capak$^{2,3}$ \\
$^{1}${Dark Cosmology Centre, Niels Bohr Institute, Juliane Maries Vej 30, 2100 K\o benhavn, Denmark} \\
$^{2}${California Institute of Technology, MC 105-24, 1200 East California Blvd., Pasadena, CA 91125, USA} \\
$^{3}${Infrared Processing and Analysis Center, California Institute of Technology, MC 100-22, 770 South Wilson Ave., Pasadena, CA 91125, USA}}

\date{Accepted XXX. Received YYY; in original form ZZZ}

\pubyear{2016}

\begin{document}
\label{firstpage}
\pagerange{\pageref{firstpage}--\pageref{lastpage}}
\maketitle

% Abstract of the paper
\begin{abstract}
We combine star formation along the `main sequence', quiescence, and clustering and merging to produce an empirical model for the evolution of individual galaxies.  Main sequence star formation alone would significantly steepen the stellar mass function towards low redshift, in sharp conflict with observation.  However, a combination of star formation and merging produces a consistent result for correct choice of the merger rate function.  As a result, we are motivated to propose a model in which hierarchical merging is disconnected from environmentally-independent star formation.  This model can be tested via correlation functions and would produce new constraints on clustering and merging.
\end{abstract}

% Select between one and six entries from the list of approved keywords.
% Don't make up new ones.
\begin{keywords}
galaxies: evolution, galaxies: star formation
\end{keywords}

\section{Introduction}

A key discovery over the past decade has been the development of the star-forming `main sequence'.  
Almost all star-forming galaxies at any fixed redshift $z < 6$ are observed to have a tight correlation between their star formation rates (SFR) and their existing stellar masses (cf. \citet{Noeske2007,Elbaz2007,Daddi2007,Peng2010,Tasca2015}).  Studies using a variety of selection criteria, SFR, and stellar mass ($M_*$) indicators both show strong agreement at a common redshift and are well fit with a common exponential decline in SFR at different choices of fixed $M_*$ as a function of time \citep{Speagle2014}.  There is now a strong consensus understanding of the rate at which typical star-forming galaxies make new stars at nearly every redshift where star-forming galaxies are observed.

The more carefully the implications of the star-forming main sequence are considered, the more surprising it appears to be.  Although it has been robustly measured, it appears to conflict with theoretical expectations and other observations in two key ways:
\begin{itemize}
	\item{{\bf Lack of Environmental Dependence or Individuality:} Star-forming main sequence studies typically cannot resolve individual galaxies or determine their environment.  However, color-based selection of star-forming galaxies should include all types of galaxies, and thus sample a wide range of environments.  After all, galaxies are observed to form in environments ranging from the centers of large clusters \citep{Abell1989} to field galaxies \citep{vanDokkum2005} and even in near-voids \citep{Szomoru1996}.  
		
		This environment is observed to be an important factor in determining merger rates \citep{Fakhouri2009} and AGN activity \citep{Satyapal2014,Khabiboulline2014}.  Numerical simulations also find that environment should be an important driver of star formation \citep{Hirschmann2014,Genel2016}.  The importance of environment is also underscored by a correlation between star formation in central and satellite galaxies, an effect that has been termed `galactic conformity' \citep{Weinmann2006,Hartley2015,Kawinwanichakij2016}.  
		
		However, the narrowness of the star-forming main sequence indicates that for a star-forming galaxy, the SFR can be determined nearly exclusively by the stellar mass and cosmic epoch, with all other factors having minimal impact.  To within the $\sim 0.2$ dex scatter of the main sequence, it is not necessary to know the environment in order to determine the star-formation rate.  Nor is it necessary to know the morphology, metallicity, star-formation history, age of the stellar population, etc.  The main sequence instead indicates that even though individual galaxies end up being unique, their star-formation obeys a universal law independent of local conditions.}		 
	\item{{\bf Are Large or Small Galaxies More Efficient?} The slope of the star-forming main sequence is less than unity, so that more massive galaxies have higher SFR but lower SFR per unit mass (specific star formation rate, or sSFR).  Thus, more massive galaxies are less efficient at star formation, and would take longer to form their existing stellar mass if it were all formed on the main sequence.  Recent work suggests that the SFR-$M_*$ relation may further flatten at high stellar masses, resulting in large galaxies being even less efficient per unit mass than previously believed \citep{Whitaker2014,Lee2015,Tomczak2016}.
		
	  However, a variety of other observations instead find that more massive galaxies are more efficient.  Although more massive halos virialize later than less massive ones \citep{Press1974,Sheth2001,Springel2005,Vogelsberger2014}), the most massive galaxies appear to finish their star formation earlier than less massive ones, an effect often termed mass `downsizing' (cf. \citet{Cowie1996}).   Further, in the mass regime most commonly probed by the star-forming main sequence ($\sim 10^{8-10} M_\odot$), at higher halo masses a greater fraction of the baryonic mass is processed into stars, with efficiency peaking around a halo mass of $10^{12} M_\odot$ \citep{Leauthaud2012b,Gonzalez2013}.  Finally, both ULTRAVISTA \citep{Ilbert2013} and ZFOURGE \cite{Tomczak2014} find that the low-mass slope of the stellar mass function flattens towards low redshift, with ULTRAVISTA reporting a sharper effect than ZFOURGE}.  However, if low-mass galaxies grow more quickly than high-mass galaxies, this will instead produce a steepening slope \citep{Peng2014b}.
\end{itemize}

As a result, we are motivated to search for ways to reconcile the main sequence with the evolution of the mass function.  In this work, we build an empirical model based upon observed galactic stellar mass distributions over a wide range of redshifts.  Where possible, we have picked the simplest possible prescriptions drawn directly from observational results.  To some extent, this goes against the trend of recent modeling \citep{Conroy2009,Leja2015}, in which the goal has been to build an increasingly detailed picture of the important physics that drive star formation and galaxy evolution.  Our work, by contrast, seeks to find a minimal model consistent with the evolution of the low-mass end of the observed stellar mass function.  We ultimately produce a model including star formation along the star-forming main sequence, turnoff constrained to match observed quiescent populations, and mergers as independent, history-free events.

In \S~\ref{sec:sfms}, we demonstrate that, as predicted by \citet{Peng2014b}, the observed evolution along the star-forming main sequence alone cannot reproduce observed mass functions.  In \S~\ref{sec:quiescence}, we demonstrate that the observed quenching of some galaxies towards low redshift cannot resolve this discrepancy.  

\citet{Peng2014b} suggested that an appropriate choice of merger rate function might fix the problem, but had insufficient data to determine whether the required merger parameters would be physically reasonable.  In \S~\ref{sec:mergers}, we show that the correct choice of merger rate function, combined with the star forming main sequence, can indeed reproduce the observed evolution in galactic mass functions.  Further, this merger rate function is also supported by numerical simulations.  As a result, we are able to produce a new model, described in \S~\ref{sec:discussion}, in which the star-forming main sequence and mergers combine to match observed galaxy distributions over a wide range of redshift.  This model also makes specific predictions for clustering and merging parameters that are currently poorly constrained by existing observations.

This work uses a $(h, \Omega_m, \Omega_\Lambda) = (0.704, 0.272, 0.728)$ cosmology \citep{Planck2015} throughout.  

\section{The Star-Forming Main Sequence and Mass Functions}
\label{sec:sfms}
It is now the observational consensus that almost all star-forming galaxies lie on a 'main sequence', linking stellar mass, star formation rate, and redshift at least out to {$z=6$} \citep{Noeske2007,Duncan2014,Steinhardt2014a}. This relation has a relatively low scatter ($\sigma\approx0.2$ dex at all redshifts; \citet{Speagle2014}).  It would therefore appear that most galaxies at a common mass formed their stars at a similar rate and time, and thus share a common history of star formation \citep{Steinhardt2014c}.  

In principle, it should be possible to constrain this common history through a continuity analysis, beginning with a measured stellar mass function at some initial cosmic epoch $\tau_{i}$ and requiring that the net effects from star-formation, aging of the existing stellar population, and merging over a given period of time $\Delta\tau$ combine to produce the measured stellar mass function at some final $\tau_f$ (cf. \citet{Ilbert2013}).  

{Previous work along these lines \citep{Behroozi2013,Peng2014b,Tomczak2016,Contini2016} has resulted in the development of increasingly complex models in order to attempt to match the observed evolution of the mass function.  For example, \citet{Tomczak2016} found that star formation and merging along required an unphysically high merger rate, whereas \citet{Contini2016} also include stellar stripping in an attempt to better match stellar mass functions.  The many possible parameters and complex feedback mechanisms involved in galactic evolution result in models very easily becoming underconstrained by observation.  Here, we search for the minimal model that is consistent with observed mass functions.  

A natural first attempt is to consider whether stellar mass growth in typical galaxies could be dominated by evolution along the star-forming main sequence, with negligible influence from mergers and other environmental factors.  After all, at any given time, only a small fraction of high-redshift galaxies are undergoing a major merger \citep{Leauthaud2012a}, and at redshifts $z > 1$ relatively few galaxies with {$\log(M_*)<10.5$} are observed to be quiescent \citep{Ilbert2013}.  The remainder will lie on the star-forming main sequence during the period when they apparently form most of their stars.  

Qualitatively, observed stellar mass functions are characterized by two properties, each of which we must be able to match: (1) they are reasonably approximated by Schechter functions 
\begin{equation}
n(M)dM = \phi e^{-M/M^*}(M/M^*)^\alpha\frac{dM}{M},
\end{equation}
 at all redshifts $z < 4$ where they have been well-measured; and (2) the low-mass slope of the Schechter function is initially steep ($\alpha \sim -1.6$), with a much larger number density of galaxies with $M_*/M_\odot \sim 10^9$ than $M_*/M_\odot \sim 10^{10}$, but flattens out gradually towards $\alpha \sim -1.4$ at lower redshifts ($z < 1.5$) \citep{Ilbert2013}.  

The principal study from which the observed star-forming, quiescent, and overall stellar mass functions used in this work are drawn \citep{Ilbert2013} fits some mass functions with a standard Schechter function and others with a double Schechter function, 
\begin{equation}
n(M)dM = e^{-M/M^*}\left[{\phi_1}(M/M^*)^{\alpha_1} + {\phi_2}(M/M^*)^{\alpha_2}\right]\frac{dM}{M}.
\end{equation}
Because the double Schechter function was adopted primarily in order to produce the bright-end shape and this work is primarily concerned with the faint-end slope, this difference should be negligible.  We have chosen to use a single Schechter function at all redshifts for consistency, re-fitting the mass functions in \citet{Ilbert2013} to produce new parameters where required (Table \ref{tab:schechter}).  For the remainder of this work, we describe observed mass functions in terms of the three parameters of their best-fit Schechter functions: a normalization $\phi$, turnover mass $M_*$, and slope $\alpha$.  \begin{center}
\begin{table}
\centering
\caption{Best-fit Schechter function parameters for observed total, star-forming, and quiescent mass functions at $0.2 < z < 4.0$, adapted from Ilbert et al. (2013)}
\label{tab:schechter}
\begin{tabular}{ |c|c|c|c| } 
\hline
$z$ & log($M_*/M_\odot$) & $\phi$ ($10^{-3}$ Mpc$^{-3}$) & $\alpha$ \\
 \hline
0.2 -- 0.5 & 10.88 & 1.88 & -1.25\\
SF only & 10.60 & 2.14 & -1.23\\
Quiescent & 10.91 & 0.94 & -0.95\\
0.5 -- 0.8 &  11.03 & 0.97 & -1.35\\
SF only & 10.62 & 1.52 & -1.29\\
Quiescent & 10.93 & 1.11 & -0.46\\
0.8 -- 1.1 & 10.87 & 1.33 & -1.32\\
SF only & 10.80 & 0.82 & -1.40\\
Quiescent & 10.81 & 1.57 & -0.11\\
1.1 -- 1.5 & 10.71 & 1.56 & -1.27\\
SF only & 10.67 & 1.31 & -1.27\\
Quiescent & 10.72 & 0.70 & 0.04\\
1.5 -- 2.0 & 10.74 & 0.86 & -1.39\\
SF only & 10.66 & 0.94 & -1.39\\
Quiescent & 10.73 & 0.22 & 0.10\\
2.0 -- 2.5 & 10.74 & 0.51 & -1.33\\
SF only & 10.73 & 0.46 & -1.35\\
Quiescent & 10.59 & 0.10 & 0.88\\
2.5 -- 3.0 & 10.76 & 0.29 & -1.43\\
SF only & 10.90 & 0.19 & -1.49\\
Quiescent & 10.27 & 0.003 & 3.26\\
3.0 -- 4.0 & 10.74 & 0.12 & -1.54\\
SF only & 10.74 & 0.12 & -1.56\\
Quiescent & & Too few & \\
 \hline
\end{tabular}
\end{table}
\end{center}

\citet{Speagle2014} find that the slope of the star-forming main sequence is well fit by $\textrm{SFR}/M_* \sim M_*^ {-0.16-0.026t}$ since at least {$z \sim 4$} (where {$t$} is the age of the universe in GYr), indicating that larger galaxies have lower specific star formation rates (sSFR $= \textrm{SFR}/M_*$). In \citet{Peng2014b} it was shown that main sequence laws of the form {$\textrm{sSFR} \sim M_*^{-\beta}$} for some positive constant {$\beta$} produce steepening in the faint end slope of the galactic mass function over time.  Their analytical results do not formally hold for a time-varying {$\beta$}, but the fact that {$\beta(t) = -0.16-0.026t$} is always negative indicates that a similar overall steepening should occur for evolution along our updated main sequence.  Recent evidence that the high-mass end of the star-forming main sequence may be turning over \citep{Lee2015} would provide a larger $\beta$ and an even sharper steepening above $M_*$, although this work focuses on the low-mass end.

\subsection{Simulated Evolution}

In order to investigate this, galaxy populations drawn from higher-redshift observed stellar mass functions were evolved along the central values of the star-forming main sequence, then compared with observations at lower redshifts.  The goal of these simulations was to develop the simplest model consistent with observed mass functions.  Therefore, the initial scenario included only main sequence star formation, since this is most directly observed.  In following sections we also include models for quiescence (\S~\ref{sec:quiescence}) and mergers (\S~\ref{sec:mergers}) in order to produce a better match with observation.

Galaxies drawn from the observed $z = 0.9$ mass function \citep{Ilbert2013} were evolved along the star-forming main sequence until $z = 0.3$, with the assumption that they remained on the main sequence for that entire time (Fig. \ref{fig:sfmsonly}). 
\begin{figure}
%\plotone{ms_ilbert_mf.ps}
\includegraphics[width=\columnwidth]{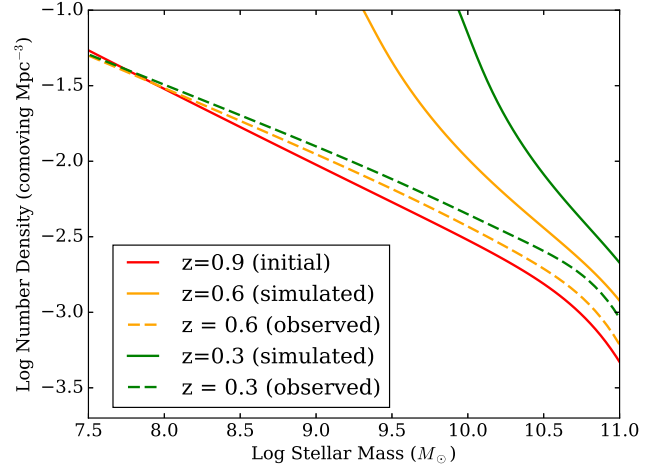}
\caption{Simulated stellar mass function evolution from $z = 0.9$ (red) to $z = 0.3$ (green) along the star-forming main sequence, beginning with the observed stellar mass function at {$z=.9$}. This main sequence evolution produces a sharp increase in the faint end slope of the best-fit Schechter function, from $-1.48$ at $z=.9$ to $-2.04$ at $z=.3$ (solid).  Observed mass functions instead have shallower slopes (dashed).}
\label{fig:sfmsonly}
\end{figure}
The resulting mass functions have a far different functional form than the observed mass function.  If approximated with a Schechter function despite the poor fit, these populations increase in faint-end slope from -1.48 ($z = 0.9$) to -2.04 ($z = 0.3$).  However, the observed stellar mass function instead becomes shallower, with a best-fit slope of $-1.40$ at $z = 0.3$.  

Since the exponent in the star-forming main sequence sSFR is always negative, this steepening would be even sharper if the simulation had been started from higher redshifts.  However, the faint end of the observed stellar mass function instead becomes slightly flatter towards low redshift.  This effect can be reduced if galaxies have a low duty cycle for star formation, but as long as the time-averaged sSFR is lower for more massive galaxies, the stellar mass function will steepen towards low redshift.  Moreover, the total stellar mass produced far exceeds that observed at low redshift, so that even if the slope were correct, the number density of galaxies would be too high at all masses.  We therefore conclude that, as suggested by \citet{Peng2014b}, evolution along the main sequence alone cannot account for the observed evolution in the mass function.  

\subsection{Computational Limitations}

In principle, it would be best to track a large population of galaxies, evolving each independently along the star-forming main sequence.  However, the computational complexity and memory requirements for such a study would have severely limited the space of merger models considered in \S~\ref{sec:mergers}. As a result, rather than evolving individual galaxies, the simulation instead evolves the mass function as a whole, dividing it into small bins and tracking the number density and its evolution in each bin.  The key problem is that mergers between galaxies at very different masses are both common and important to the mass function evolution, so galaxies over a very broad mass range must all be considered as part of the same simulation.  However, the number density of galaxies changes by many orders of magnitude over such a mass range.  Thus, tracking a statistically significant sample of rare, high-mass galaxies would have required simultaneously tracking too many accompanying low-mass galaxies to be computationally feasible.  

Such a strategy requires a more careful consideration of mathematical precision, rounding, and binning errors than tracking individual galaxies.  In addition to verifying these choices were theoretically unbiased, our results were compared with a simulation tracking smaller numbers of individual galaxies in order to confirm their validity.  Each simulation was primarily analyzed over $z = 1.7$ to $z = 0.3$, the range for which both the total and quiescent mass functions are well constrained in \citet{Ilbert2013}, with additional but lower-quality constraints available from $z = 2.7$ to $z = 1.7$.  Because small errors in mass functions build up over the course of the simulation, our results are primarily presented in intervals from $z = 1.7$ to $z = 0.9$ or $z = 0.9$ to $z = 0.3$, with both ranges showing similar behavior.

\section{Quiescence}
\label{sec:quiescence} 
Although nearly all galaxies are star forming at the highest redshifts \citep{Bouwens2015,Steinhardt2014a}, at lower redshifts there is an increasingly numerous quiescent population.  There are two main ways in {\bf which} galaxies might appear quiescent; either (1) each galaxy goes through a continuous period of star formation for some length of time and then enters a permanent state of quiesence or (2) each galaxy goes through alternating periods of star formation and quiescence.  Both would reduce the total stellar mass produced, albeit in different ways.

\subsection{Permanent Turnoff}
\label{subsec:qturnoff}
The simplest way to model permanent turnoff is to choose a quiescent mass function $n_q(M_*,t)$ as a function of cosmic epoch, then at each timestep, remove galaxies from the star-forming population accordingly, freezing their stellar masses.  For $n_q(M_*,t) = 0$, all galaxies grow continuously along the star-forming main sequence as in \S~\ref{sec:sfms}.  

Mathematically, it must always be possible to choose $n_q$ such that the resulting simulated mass functions match observed mass functions.  To find such $n_q$, after each time step the observed star-forming mass function at the new redshift was interpolated from binned mass functions and the correct number density of galaxies was pulled out of the star-forming population to match that interpolation. All galaxies pulled out of the star-forming population were assumed to have become permanently quiescent.

Beginning with the observed star-forming and quiescent mass functions at $z = 1.7$ \citep{Ilbert2013}, it is therefore possible to find $n_q(M_*,t)$ such that the total mass functions at $z = 0.9$ and $z = 0.3$ are matched by this model (Fig. \ref{fig:qusim}).
\begin{figure}
%\plotone{QuSim.ps}
\includegraphics[width=\columnwidth]{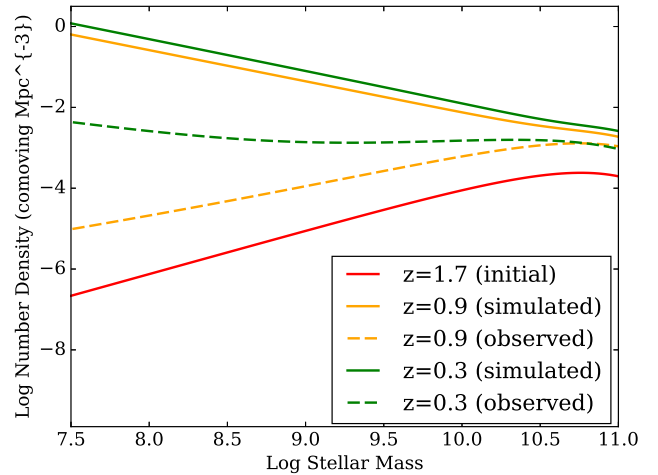}
\caption{Quiescent mass functions derived from combining main sequence growth for star-forming galaxies with permanent turnoff, constrained to match observed mass functions from \citet{Ilbert2013} at $1.7 < z < 0.3$.  The resulting quiescent mass functions $n_q(M_*,t)$ at $z = 0.9$ (yellow) and $z = 0.3$ (green) are compared with observed quiescent mass functions given in \citet{Ilbert2013}.  Although mathematically the observed total mass function can be matched at all redshifts, doing so predicts a far larger quiescent population than observed at lower redshifts.}
\label{fig:qusim}
\end{figure}
Following this path, the predicted quiescent mass functions require a far more numerous population than observed at lower redshifts.  This indiciates that permanent galactic quiescence does not solve the high-mass overproduction problem created by main sequence star formation.

\subsection{Duty Cycles}
\label{subsec:qdc}
We now consider whether lower-mass galaxies might only become temporarily quiescent rather than permanently turning off, alternating between star-forming and quiescent periods.  If star formation has a duty cycle of, e.g., 60\%, there would be 60\% as much stellar mass growth as implied by the star-forming main sequence.  Because quiescence would be temporary, it might be hoped that this could reduce the growth of low-mass galaxies without the permanent turnoff that overproduces quiescent, low-redshift galaxies.  Applying a mass-independent duty cycle would not change the exponent in the main sequence SSFR and therefore would not fix the steepening faint-end slope. Thus, in order to match the observed star-forming mass function, the duty cycle must be mass-dependent.

It should also be noted that studies of the star-forming main sequence do not measure an instantaneous SFR, but rather attempt to estimate the number of luminous, high-mass, blue stars with lifetimes $\sim 10^7$ yr that dominate the spectral energy distributions of young stellar populations.  The main sequence thus describes average star formation rates over $\sim 10^7-10^8$ yr in star-forming galaxies \citep{Kennicutt2012,Hayward2014}.  As a result, the periods of quiescence or star formation in this model would need to occur on longer timescales, and galaxies temporarily in the `off' state when observed would be selected as part of the quiescent population or green valley at that redshift, even though at lower redshifts they might again be star-forming.

Following a similar procedure to \S~\ref{subsec:qturnoff}, our simulation begins with the observed mass function at $z = 1.7$ and evolves it forward, assigning each galactic mass bin the duty cycle required to match the observed evolution.  This would require a duty cycle as low as 10\%, so that 90\% of star-forming galaxies would actually be selected as quiescent at $1.7 < z < 0.3$ (Fig. \ref{fig:dutycycles}), in addition to 100\% of post-turnoff galaxies.
\begin{figure}
\includegraphics[width=\columnwidth]{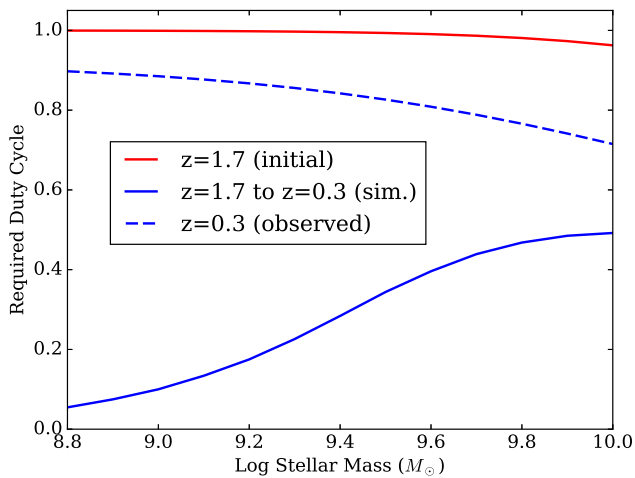}
\caption{Average duty cycle (blue, solid) for the star formation that would be required as a function of mass in order to evolve the mass function from $z = 1.7$ to $z = 0.3$ consistent with observations.  Low-mass galaxies must have lower duty cycles in order to flatten the slope of the resulting function.  Although this quiescence is transitory, these low duty cycles still predict large low-mass quiescent populations at these redshifts.  These would be far larger than the observed quiescent fraction of galaxies (\citet{Ilbert2013}; red, blue dashed), particularly on the low-mass end.}
\label{fig:dutycycles}
\end{figure} 

However, at these redshifts, at most 40\% of all galaxies with {$\log M_*<10$} are selected as quiescent \citep{Ilbert2013}.   Since neither quiescent scenario can come close to producing the correct evolution the galactic mass function, the solution must lie elsewhere.

\section{Mergers as a Solution}
\label{sec:mergers}
Finally, we consider the possibility that mergers might flatten the slope of the mass function.  This would be a surprising answer because the star-forming main sequence implies that stellar mass growth is similar on cosmic scales, yet mergers are inherently environmental and local.  Mergers also would flatten the mass function via a characteristically mechanism: rather than reducing the stellar mass growth in low-mass galaxies, those galaxies grow quickly but then disappear, absorbed into more massive ones.

Although major mergers between two galaxies of similar mass are relatively rare, simulations suggest that the absorption of small galaxies by much larger galaxies is fairly common \citep{Fakhouri2010}. The merger rate between dark matter haloes in the Millennium simulation is well fit by the functional form
\begin{equation}
%\begin{align*}
\label{eq:f10}
  \frac{\dif N_m}{\dif\xi\dif z}(M,\xi,z) = A M^\alpha\xi^\beta\exp\left[
  \left(\frac{\xi}{\tilde\xi}\right)^\gamma\right](1+z)^\eta,
%\end{align*}
\end{equation}
where {$N_m$} is the number density of mergers, {$M$} is the mass of the larger galaxy, {$\xi$} is the mass ratio between the smaller and larger galaxy, $A$ is a normalization constant, $z$ is redshift, and $\alpha$, $\beta$, $\gamma$, and $\eta$ are parameters determined by the physical model.  The stellar mass-halo mass relation from \citet{Behroozi2013} was used to determine the effects of these mergers on stellar mass functions.  As with star formation, the simulation calculates the number density of mergers and the resulting effect on the binned mass function in small time steps, stepping through the desired redshift range.

Small galaxies are indeed absorbed in mergers far more often than their more massive counterparts.  Thus, in the absence of star formation, mergers alone will  result in a shallower low-mass mass function slope towards lower redshift (Fig. \ref{fig:mergersonly}).  
\begin{figure}
%\plotone{MeSim.ps}
\includegraphics[width=\columnwidth]{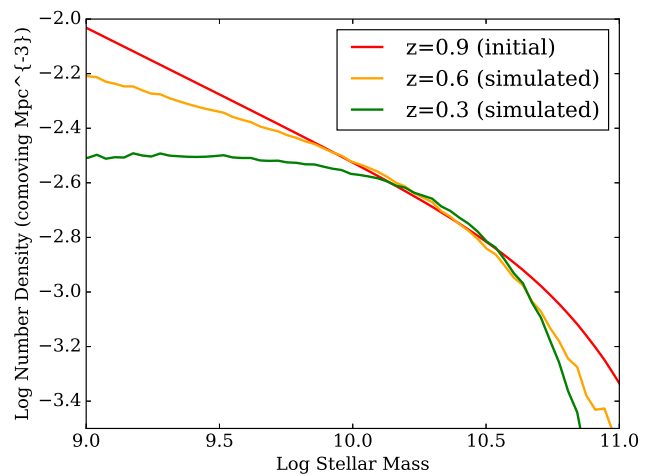}
\caption{Simulation considering the effects of mergers in isolation, beginning with the observed $z = 0.9$ mass function and evolving forward to $z = 0.3$.  The low-mass slope becomes flatter, potentially counteracting the effects of star formation.  Although it may appear mass is not conserved, the simulation finds that these seemingly `missing' galaxies end up in massive clusters, well beyond the boundaries of this figure.}
\label{fig:mergersonly}
\end{figure} 
This acts in the opposite direction of the steepening due to evolution along the star-forming main sequence.

\subsection{Matching Observed Mass Functions}
\label{subsec:merger_match}
Because dark matter simulations provide insufficient constraints on merger rates for our purposes, instead we adopt a similar approach to our analysis of quiescent populations.  The galaxy population is divided into star-forming and quiescent populations, with quiescent populations constrained to match observed quiescent mass functions.  Then, a merger rate function is selected, the galaxy population is evolved forward in time, and the final mass function compard with observations.

Using the merger rate functional form from \citet{Fakhouri2010} with the proper slope and normalization parameters chosen from Fig. \ref{fig:overlap} to evolve galaxies from $z = 1.7$ to $z = 0.9$ produces a promising result (Fig. \ref{fig:f10merging}).  Fitting the $z=0.9$ simulated curve to the double Schechter form from \citet{Ilbert2013} at $z=0.8-1.1$ with a fixed $\mathcal{M}^*=10.87$ yields $(\Phi_1^*,\alpha_1,\Phi_2^*,\alpha_2) = (2.09, -0.57, 0.39, -1.53)$.  This corresponds to an average $\chi^2$ of 0.42 across the four parameters, which indicates possible consistency although it is difficult to properly account for the degrees of freedom allowed by our model, as the parameters are partially degenerate.

\begin{figure}
\includegraphics[width=\columnwidth]{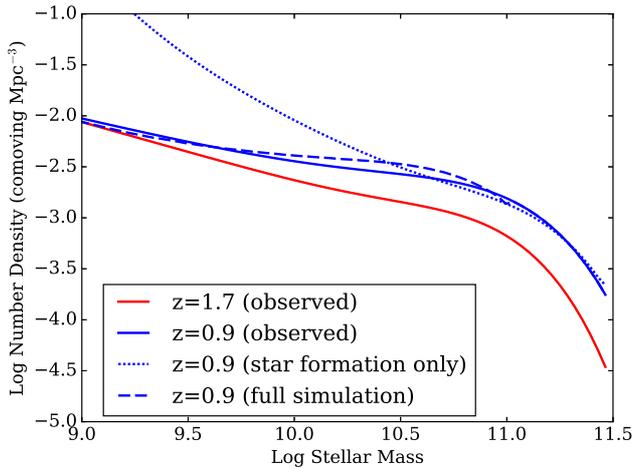}
\caption{Simulation of galaxies evolving only with main sequence star-formation (dotted), and a combination of main sequence star formation, quiesence, and mergers (dashed) from $z = 1.7$ to $z = 0.9$. This uses total mass functions from \citet{Ilbert2013} and the merger rate function form from \citet{Fakhouri2010} with well-chosen parameters. In both evolution cases parameters were picked to match the total mass function at the high-mass end {\bf ($\log M > 11$)}. With star formation only we get far too many low-mass galaxies, but with mergers these are absorbed out of the population and we get a close match with observations.}
\label{fig:f10merging}
\end{figure}

The merger rate function produced by simulations depends upon complex physics, and varying those physical parameters is beyond the scope of this paper.  Instead, we elect to describe possible merger rate functions in terms of the functional form given in (\ref{eq:f10}), producing two parameters: a slope and normalization.  A grid of possibilities is then evaluated to determine which are consistent with observations.

Our simulation is run over three redshift ranges: $2.7 < z < 1.7$, $1.7 < z < 0.9$, and $0.9 < z < 0.3$.  In each case, some parameters are indeed consistent with the observed mass function at the final redshift (Fig. \ref{fig:overlap}).  The complex shape of the allowed parameter space is primarily due to differences on the high-mass end resulting from the observed mass function having been approximated with a Schechter function at both the initial and final redshifts.
\begin{figure*}
\includegraphics[width=\textwidth]{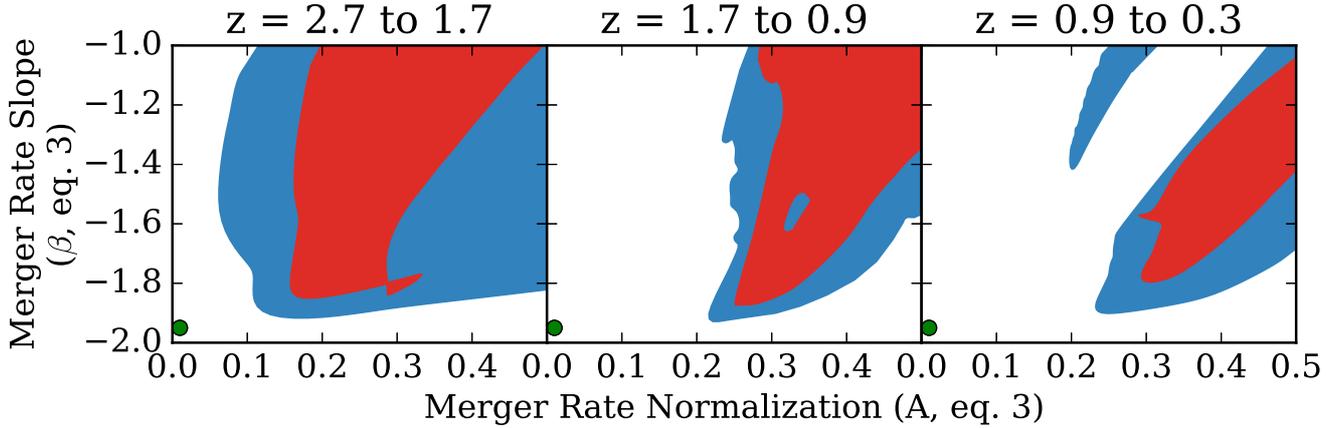}
\caption{Choices of normalization ($A$, Eq. \ref{eq:f10}) and slope ($\beta$, Eq. \ref{eq:f10}) for merger rate functions producing mass functions consistent with observations at the $1\sigma$ (red) and $2\sigma$ (blue) levels for simulations run over three redshift ranges: $2.7 < z < 1.7$ (left), $1.7 < z < 0.9$ (center), and $0.9 < z < 0.3$ (right).  There is some overlap between these regions, so a common merger rate functions would suffice over all three redshift ranges.  The parameters corresponding to \citet{Fakhouri2010} are shown on each panel in green.}
\label{fig:overlap}
\end{figure*}
For some choices of parameters, similar merger rate functions are consistent with observations over all three redshift ranges.  The merger rate function given by \citet{Fakhouri2010} does not lie in this allowed space.  However, similar merger rate functions would be plausible.  

\subsection{Comparing Observations with Quantitative Merger Simulations}
\label{subsec:mergerdiff}

It should be noted that the merger rate functions produced here describe luminous galaxies rather than the dark matter halos described by numerical simulations of hierarchical merging.  Thus, although it is tempting to interpret the slope and magnitude of the merger rate functions in Fig. \ref{fig:f10merging} as candidates for dark matter halo merger rates, there are several reasons to be wary of a direct comparison.  A proper translation requires not just matching halo masses with stellar masses, but also understanding how halo occupation rates vary between halos of the same mass in rich environments with frequent mergers and in sparse environments or even voids.

If all halos have a constant occupation rate, independent both of mass and of the likelihood that they will merge, then this effect is negligible.  For example, if 50\% of halos have galaxies, there are twice as many halos as calculated from galaxy mass function at every mass, and thus four times as many mergers between any pair of masses.  However, only 1/4 of those mergers will involve occupied halos, producing the same rate for merging a pair of galaxies at those masses.  

However, it is very likely that halos in dense environments are more likely to be occupied, and thus the translation between halo merger rates and their effect on galaxy mass functions is more complex.  In that regard, the merger rate functions described here are degenerate combinations of halo merger rates and occupation fraction.  They describe merger rates between luminous galaxies, and care must be taken when comparing them with predicted merger rates for dark matter halos.

The rate at which a relatively small galaxy ({$10^9$} solar masses in Fig. \ref{fig:absorb}) is absorbed into larger galaxies of various masses is strongly mass-dependent (Eq. \ref{eq:f10}). 
\begin{figure}
\includegraphics[width=\columnwidth]{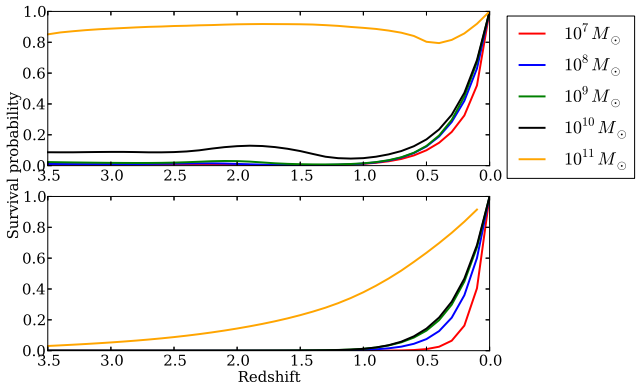}
\caption{The probability that (top) star-forming and (bottom) quiescent galaxies will survive (not be absorbed by a larger galaxy via merger) as a function of mass and redshift, according to prescription given in \citet{Fakhouri2010}.  Note that these probabilities are dominated by rare, high-mass galaxies, making them difficult to constrain.}
\label{fig:absorb}
\end{figure} 
Given current mass functions, we find that most small galaxies involved in mergers are absorbed by large ({$\log M_*>11$}) galaxies, even though there are relatively few large galaxies in the population. As a result, the overall merger rate of small galaxies is very sensitive to the total number of large galaxies.

Surprisingly, even relatively high-mass galaxies are likely to be absorbed by a larger object.  For example, a $M_* = 10^{10} M_\odot$ star-forming galaxy at redshift 3.5, despite being one of the largest galaxies in existence at that time, has only a 9\% probability of not merging with a larger object by redshift 0.  For these galaxies, the larger object is likely the formation of a massive galaxy cluster, because individual galaxies are not seen with, e.g., $M_* \sim 10^{14} M_\odot$.  Thus, these galaxies will often indeed survive to redshift 0, although many studies of mass functions exclude them and focus on field galaxies (cf. \citet{Bundy2015}).

The uncertainty in the observed galactic mass function is relatively large at the high mass end due to the rarity of high-mass galaxies. There is a corresponding increase in uncertainty at the high mass end of the stellar mass-halo mass relation (particularly at high redshifts), making it hard to match high stellar mass galaxies to their simulated halo merger rates.  Varying these parameters does not significantly affect the overall shape of the merger rate function, so the flattening effect that mergers have on the overall mass function is fairly robust qualitatively. However, the slope and magnitude of this merger rate curve, and therefore the importance of mergers in determining mass functions, cannot be well constrained from halo simulations.

\section{Discussion}
\label{sec:discussion}

The star-forming main sequence appears to exist at all redshifts where it can be measured.  However, its tightness and universality would appear to disagree with both theoretical expectations and observational evidence that environment and merger histories are also key drivers of galaxy evolution.  Further, a continuity analysis shows that growth along the main sequence alone cannot reproduce observed stellar mass functions.  In this work, we propose a possible solution: the star-forming main sequence is indeed a good description of star formation, but merging is also necessary to describe the growth of stellar mass.  Counterintuitively, postulating a complete separation between the process of acquiring the ingredients that will become stars (clustering and merging) and the process of turning that into new stars (main sequence star formation) can produce a model consistent with both theoretical and observational constraints.

We also note that a consistent model can be developed in which the duty cycle for star formation is 100\%.  Galaxies would spend one long, extended period forming new stars along the main sequence followed by permanent quiescence.  This is consistent with predictions made via measurements of synchronization timescales along the main sequence and with quasars \citep{Steinhardt2014c}.

The next step is to understand whether this empirical model can be described in terms of fundamental astrophysical processes.  Merging is responsible for the acquisition of hot gas, yet star formation requires further cooling (cf. \citet{Bromm2013}).  Perhaps, the rate limiting step lies in cooling channels, and as long as there is a surplus of hot gas available, star formation will be similar whether that hot gas was added in a recent merger or in the distant past.  In that case, the time-dependence of the star-forming main sequence may be due to declining background galaxy temperatures, perhaps associated with a cooling cosmic microwave background.  Strong correlations between SFR and H II regions yet weaker correlations with H I (cf. \citet{Bigiel2008}) may hint at a similar outcome.  

This model also presents requirements for the merger rate function, reducing a two-dimensional space of possible parameters to a one-dimensional locus.  Current simulations are underconstrained at high-redshift, so this produces a useful new set of constraints.

These constraints are primarily produced through a focus on the slope of the low-mass end of the observed stellar mass function.  Several additional effects become important on the high-mass end, most notably stellar stripping \citep{Murante2007,Purcell2007,Guo2012,Contini2014}.  Properly modeling stellar stripping requires tracking additional history and is dependent upon environment, while one of the surprising features of the model in this paper is that the low-mass end can be matched with a history-free model, treating the various processes involved as independent and random.  Since stellar stripping is negligible below the `knee' of the stellar mass function \citep{Contini2016}, it potentially provides a mechanism for improving the fit between predicted and observed mass functions through independently correcting for differences on the high-mass end.

Finally, we note that this model potentially provides a mechanism for explaining another longstanding puzzle, one of several effects often termed downsizing \citep{Cowie1996}.  It is observed that typical massive galaxies finish their formation earlier than lower-mass galaxies.  However, smaller halos should form earlier \citep{Sheth2001,Springel2005,Vogelsberger2014}, and the main sequence similarly indicates that low-mass galaxies form more efficiently.  Considerable work has therefore gone into searching for models in which early, massive galaxies might become more efficient (e.g., \citet{Somerville2015}).  Our new model proposes an entirely different solution: low-mass galaxies indeed form efficiently at high redshift, but they have a high probability of being absorbed into more massive ones.  

The new empirical model presented in this work suggests several promising possibilities for unifying several different processes involved in galaxy evolution.  If these behaviors can be produced via reasonable astrophysics, it could provide a solution to several key problems, most notably the remarkable disconnect between the star-forming main sequence and the remainder of our knowledge about how galaxies grow.  However, this requires that such a model can be generated from meaningful cosmology and astrophysics, which requires considerable further analysis.

The authors would like to thank Micaela Bagley, Andrew Benson, Zachary Claytor, Andreas Faisst, Olivier Ilbert, Adam Jermyn, Ian Kuehne, Dan Masters, Daniel McAndrew, and Sune Toft for helpful comments.  CS acknowledges support from the ERC Consolidator Grant funding scheme (project ConTExt, grant number No. 648179) and from the Carlsberg Foundation.  DY was supported by the Victor Neher SURF Fellowship;

\bibliographystyle{mnras}

\end{document}